\documentclass[12pt]{article}

\usepackage{amsmath,amssymb,amsfonts,amsbsy}
\usepackage{cite}
\usepackage{graphicx}
\usepackage{wrapfig}
\usepackage{epsfig}

\usepackage{bbm} 
\usepackage{bm} 
\usepackage{color}                                                       %
\usepackage{dsfont} 
\usepackage{latexsym} 
\usepackage{lscape} 
\usepackage{mathrsfs} 
\usepackage{morefloats} 
\usepackage{floatflt} 
\usepackage{slashed} 
\usepackage{psfrag}

\DeclareFontFamily{OT1}{mygreek}{}%
\DeclareFontShape{OT1}{mygreek}{m}{n}{<->omsegr}{}%
\DeclareFontShape{OT1}{mygreek}{b}{n}{<->omsegrb}{}%
\DeclareFontShape{OT1}{mygreek}{m}{it}{<->omsegri}{}%
\DeclareFontShape{OT1}{mygreek}{bx}{n}{<->sub * mygreek/b/n}{}%
\DeclareFontShape{OT1}{mygreek}{m}{sl}{<->sub * mygreek/m/it}{}%
\DeclareSymbolFont{Greekrm}{OT1}{mygreek}{m}{n}
\DeclareSymbolFont{Greekbf}{OT1}{mygreek}{b}{n}
\DeclareSymbolFont{Greekit}{OT1}{mygreek}{m}{it}
\DeclareMathSymbol{\omegab}{\mathalpha}{Greekbf}{119}

\begin{document}
\addcontentsline{toc}{subsection}{{Title of the article}\\
{\it B.B. Author-Speaker}}
\graphicspath{{exper/yourname/}}

\setcounter{section}{0}
\setcounter{subsection}{0}
\setcounter{equation}{0}
\setcounter{figure}{0}
\setcounter{footnote}{0}
\setcounter{table}{0}

\begin{center}
\textbf{IMPORTANCE OF SEMI INCLUSIVE DIS PROCESSES IN DETERMINING
FRAGMENTATION FUNCTIONS \footnote{{\small This research was
supported by the JINR-Bulgaria Collaborative Grant and the RFBR
Grants (Nrs 11-01-00182, 12-02-00613 and 13-02-01005).}}}

\vspace{5mm}

E.~Leader$^{\,1}$, A.V.~Sidorov$^{\,2}$ and
D.B.~Stamenov$^{\,3}$

\vspace{5mm}

\begin{small}
  (1) \emph{Imperial College, Prince Consort Road, London SW7 2BW, England} \\
  (2) \emph{Joint Institute for Nuclear Research, 141980 Dubna, Russia} \\
  (3) \emph{Institute for Nuclear Research and Nuclear
Energy, Bulgarian Academy of Sciences,\\ Blvd. Tsarigradsko
Chaussee 72, Sofia 1784, Bulgaria} \\
\end{small}
\end{center}

\vspace{0.0mm} 

\begin{abstract}
  A NLO QCD analysis of the HERMES and COMPASS data on pion
  multiplicities is presented. Sets of pion fragmentation
  functions are extracted from fits to the data and compared with
  those obtained from other groups before these data were
  available. The consistency between HERMES and COMPASS data is
  discussed. We point out a possible inconsistency between the
  HERMES $[x, z]$ and $[Q^2, z]$ presentations of their data.
\end{abstract}

\vspace{7.2mm}

In the absence of charged current neutrino data, the experiments
on polarized inclusive deep inelastic lepton-nucleon scattering
(DIS) yield information only on the sum of quark and anti-quark
parton densities $\Delta q + \Delta \bar{q}$ and the polarized
gluon density $\Delta G$. In order to extract separately $\Delta
q$ and $\Delta \bar{q}$ other reactions are needed. One
possibility is to use the {\it polarized} semi-inclusive
lepton-nucleon processes (SIDIS) $l+ N \rightarrow l'+h+X$, where
$h$ is a detected hadron (pion, kaon, etc) in the final state. In
these processes new physical quantities appear - the fragmentation
functions $D^h_{q, \bar q}(z, Q^2)$ which describe the
fragmentation of quarks and antiquarks into hadrons. Due to the
different fragmentation of quarks and anti-quarks, the polarized
parton densities $\Delta q$ and $\Delta \bar{q}$ can be determined
separately from a combined QCD analysis of the data on inclusive
and semi-inclusive asymmetries. The key role of the fragmentation
functions (FFs) for the correct determination of sea quark parton
densities $\Delta \bar{q}$ was discussed in \cite{deltas_puzzle}.
There are different sources to extract the fragmentation functions
themselves: The semi-inclusive $e^+ \, e^-$ annihilation data,
single-inclusive production of a hadron $h$ at a high transverse
momentum $p_T$ in hadron-hadron collisions, unpolarized
semi-inclusive DIS processes. It is important to mention that the
data on hadron multiplicities in unpolarized SIDIS processes are
crucial for a reliable determination of FFs, because only then one
can separate $D_q^h(z,Q^2)$ from $D_{\bar q}^h(z,Q^2)$. Such data
have been used {\it only} by the DSS group in their global
analysis \cite{DSS}. As a result, the properties of the extracted
set of FFs significantly differ, especially in the kaon sector,
from those of the other sets of FFs \cite{other_FFs}.
Unfortunately, the new properties of the DSS FFs are based on the
{\it unpublished} HERMES'05 SIDIS data on hadron multiplicities
which are not confirmed by the final HERMES data \cite{HERMES}. It
turns out that not only the DSS FFs, but all other sets of pion
and kaon FFs are NOT in agreement with the recent HERMES and
COMPASS data \cite{COMPASS} on hadron multiplicities.

In this talk we present our results on new pion fragmentation
functions extracted from a NLO QCD fit to the HERMES and COMPASS
(the first ref. in \cite{COMPASS}) data on the pion
multiplicities. While COMPASS reports data only on a deuteron
target, HERMES presents data on both the proton and deuteron
targets.

The multiplicitiy $M_{p(d)}^{\pi}(x,Q^2,z)$ of pions using a
proton (deuteron) target are defined as the number of pions
produced, normalized to the number of DIS events, and can be
expressed in terms of the semi-inclusive cross section
$\sigma_{p(d)}^{\pi}$ and the inclusive cross section
$\sigma_{p(d)}^{DIS}$:
\begin{eqnarray}
M_{p(d)}^{\pi}(x,Q^2,z)&=&
\frac{d^3N_{p(d)}^{\pi}(x,Q^2,z)/dxdQ^2dz}{d^2N_{p(d)}^{DIS}(x,Q^2)
/dxdQ^2}=\frac{d^3\sigma_{p(d)}^{\pi}(x,Q^2,z)/dxdQ^2dz}{d^2\sigma_{p(d)}
^{DIS}(x,Q^2)/dxdQ^2}\nonumber\\
&=&\frac{(1+(1-y)^2)2xF_{1p(d)}^h(x,Q^2,z)+2(1-y)xF_{Lp(d)}^h(x,Q^2,z)}
{(1+(1-y)^2)2xF_{1p(d)}(x,Q^2)+2(1-y)F_{Lp(d)}(x,Q^2)}.
\label{M_exp_th}
\end{eqnarray}
In Eq. (\ref{M_exp_th}) $F_1^h, F_L^h$ and $F_1, F_L$ are the
semi-inclusive and the usual nucleon structure function
respectively, which are expressed in terms of the unpolarized
parton densities and the fragmentation functions ($F_1^h, F_L^h$),
and by the unpolarized parton densities ($F_1, F_L$).

Let's  start our discussion with the results of the fit to COMPASS
deuteron data. In our fit we have used the $[y, x(Q^2), z]$
presentation of the data, where $y=Q^2/2MEx$ is the fractional
energy of the virtual photon, and $M$ and $E$ are the mass of the
nucleon and the energy of the muon beam, respectively.
\begin{figure} [h]
\begin{center}
  \includegraphics[height=.40\textheight]{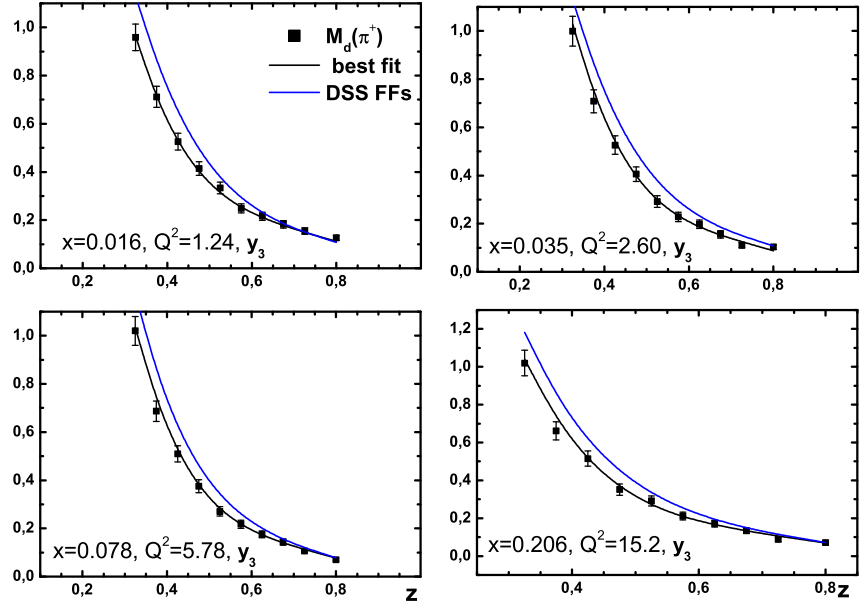}
\caption{\footnotesize Comparison of our NLO QCD results for
COMPASS $\pi^{+}$ multiplicities with the data. The multiplicities
computed with the DSS FFs are also shown.} \label{Sidorov fig1}
\end{center}
\end{figure}
The data on the multiplicities  are distributed in five y-bins as
functions of z at different fixed values of $(x, Q^2)$. The total
number of the data points is 398, 199 for $\pi^{+}$ and 199 for
$\pi^{-}$ multiplicities. The errors used are quadratic
combinations of the statistical and systematic errors. The number
of free parameters, attached to the input parametrizations of the
pion FFs [$D_u^{\pi+}(z),~D_{\bar u}^{\pi+}(z),~D_g^{\pi+}(z)$] at
$Q^2= 1~GeV^2$ and determined from the fit, is 12. The assumption
that all unfavored pion FFs are equal is used. For the unpolarized
parton densities we use the NLO MRST'02 set of PDFs \cite
{MRST02}. The charm contribution to the multiplicities is not
taken into account. For the value of $\chi^2/DOF$ corresponding to
the best fit to the data we obtain 283.12/386=0.73. An excellent
description of the COMPASS pion data is achieved. The quality of
the fit is illustrated for the $y_3$-bin [0.2-0.3] (see Fig. 1 for
$\pi^{+}$ and Fig. 2 for $\pi^{-}$ multiplicities). In the figures
are presented also the multiplicities at the COMPASS kinematics
calculated using the DSS FFs (blue curves). The extracted pion FFs
will be presented later and compared to those obtained from our
fit to the HERMES data, as well as to some of the FFs sets
available at present. Here we would like only to mention that it
is obvious that the COMPASS data are in disagreement with the DSS
FFs.
\begin{figure}
\begin{center}
  \includegraphics[height=.40\textheight]{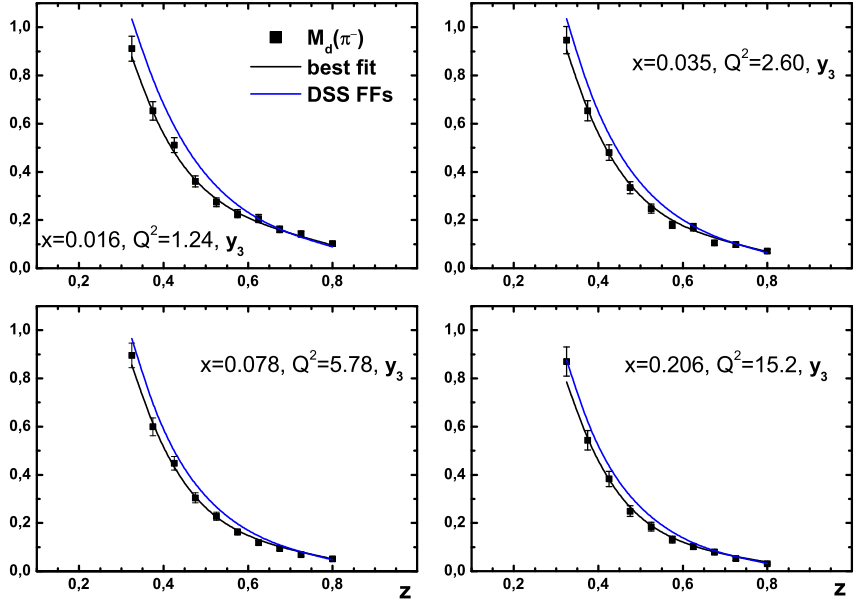}
\caption{\footnotesize Comparison of our NLO QCD results for
COMPASS $\pi^{-}$ multiplicities with the data. The multiplicities
computed with the DSS FFs are also shown.} \label{Sidorov fig2}
\end{center}
\end{figure}

Let us discuss now our results on the pion FFs extracted from a
NLO QCD fit to the HERMES proton and deuteron data on pion
multiplicities, corrected for exclusive vector meson production
\cite{HERMES}. In our analysis we have used the $[x,z]$ as well as
the $[Q^2,z]$ presentation of the data. The pion multiplicities
are given for 4 z-bins [0.2-0.3;~0.3-0.4;~0.4-0.6;~0.6-0.8] as
functions of $x$ for the $[x,z]$ or functions of $Q^2$ for the
$[Q^2,z]$ presentation. The total number of the $\pi^{+}$ and
$\pi^{-}$ data points is 144. It turned out that we can not find a
reasonable fit to the HERMES $[x,z]$ data. Also, there is a strong
indication that the HERMES $[x,z]$ and COMPASS data are not
consistent.
\begin{figure}
\begin{center}
  \includegraphics[height=.40\textheight]{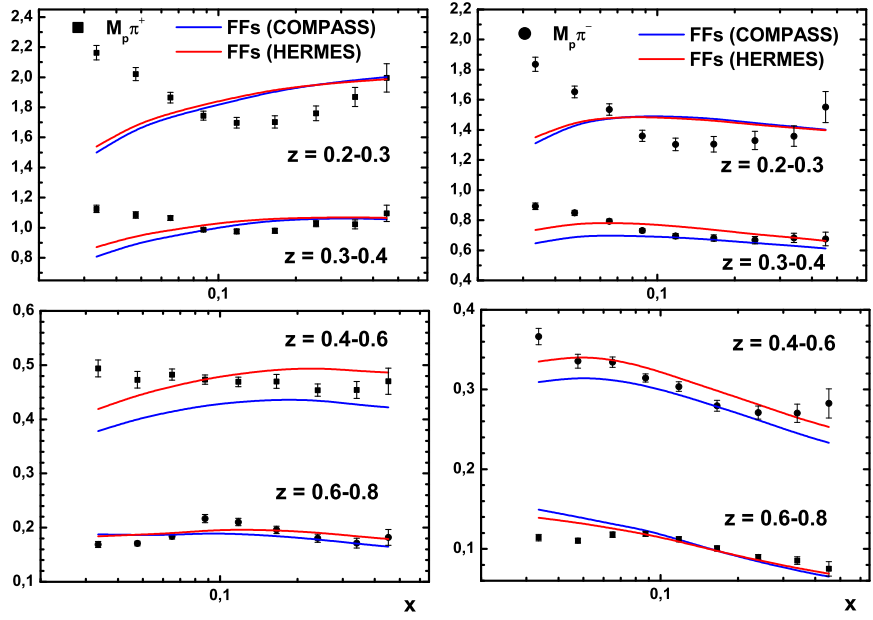}
\caption{\footnotesize Comparison of HERMES $[x,z]$ {\it proton}
data on $\pi^{+}$ (left) and $\pi^{-}$ multiplicities (right) with
the multiplicities at the same kinematic points calculated by our
FFs extracted from the COMPASS data (blue curves) and from HERMES
$[Q^2,z]$ data (red curves).} \label{Sidorov fig3}
\end{center}
\end{figure}
\begin{figure}
\begin{center}
  \includegraphics[height=.40\textheight]{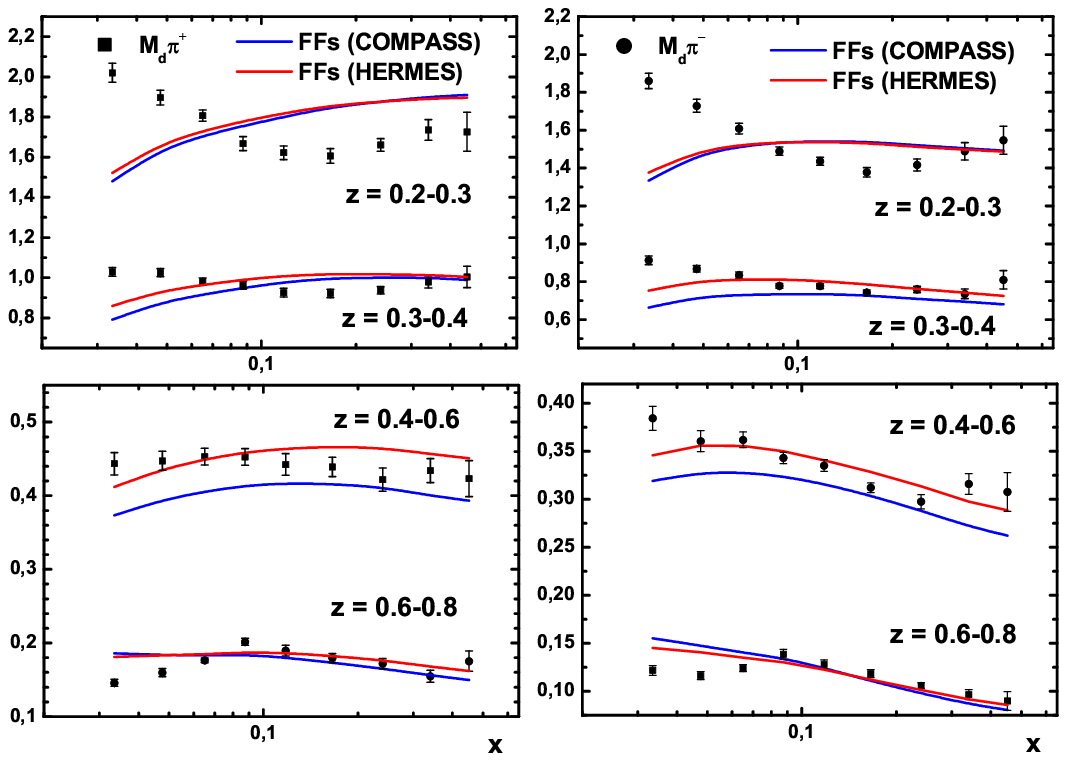}
\caption{\footnotesize Comparison of HERMES $[x,z]$ {\it deuteron}
data on $\pi^{+}$ (left) and $\pi^{-}$ multiplicities (right) with
the multiplicities at the same kinematic points calculated by our
FFs extracted from the COMPASS data (blue curves) and from HERMES
$(Q^2,z)$ data (red curves).} \label{Sidorov fig4}
\end{center}
\end{figure}
\begin{figure}
\begin{center}
  \includegraphics[height=.40\textheight]{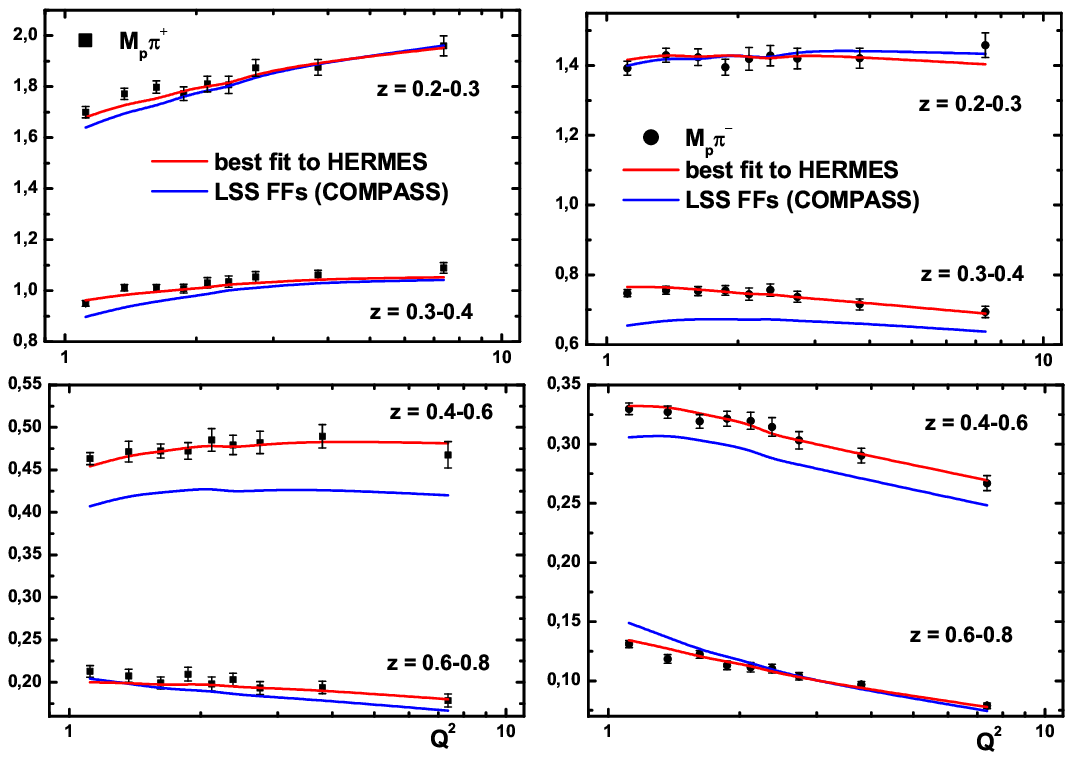}
\caption{\footnotesize Comparison of HERMES $[Q^2,z]$ {\it proton}
data on $\pi^{+}$ (left) and $\pi^{-}$ multiplicities (right) with 
the best fit
curves. The blue curves correspond to the multiplicities at the
same kinematic points calculated using our FFs extracted from the
COMPASS data.} \label{Sidorov fig5}
\end{center}
\end{figure}
\begin{figure}
\begin{center}
  \includegraphics[height=.40\textheight]{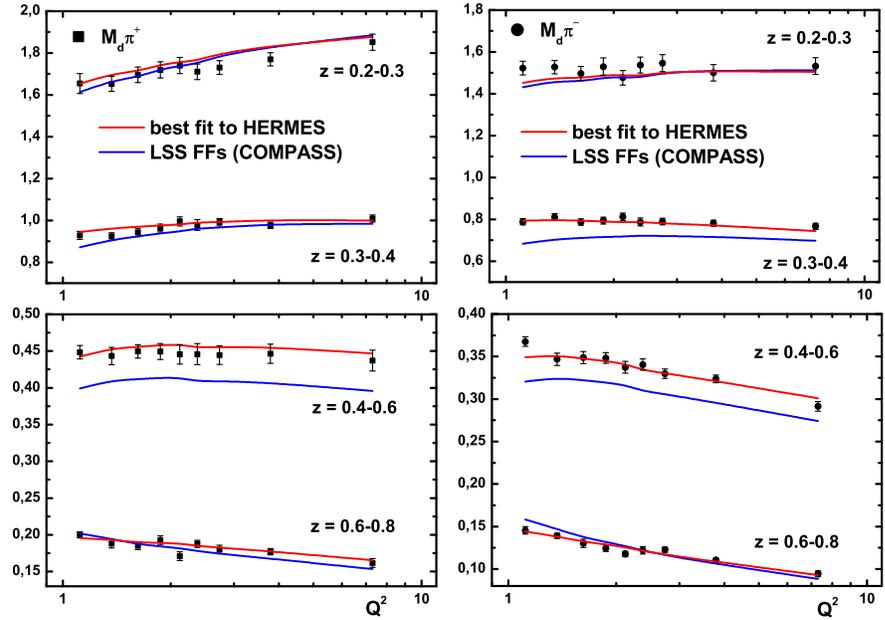}
\caption{\footnotesize Comparison of HERMES $[Q^2,z]$ {\it
deuteron} data on $\pi^{+}$ (left) and $\pi^{-}$ multiplicities
(right) with the best fit curves. The blue curves correspond to
the multiplicities at the same kinematic points calculated using
our FFs extracted from the COMPASS data.} \label{Sidorov fig6}
\end{center}
\end{figure}
We observe a big discrepancy between the values of the HERMES data
on pion multiplicities and multiplicities at the same kinematic
points computed with our FFs extracted from the COMPASS data (see
blue curves in Fig. 3 for proton and Fig. 4 for deuteron data).

We were very surprised to find that the situation is dramatically
changed if the HERMES  data on pion multiplicities in $[Q^2, z]$
presentation are used in the QCD analysis. In this case a
reasonable fit to the data is achieved, $\chi^2/DOF$ = 151.73/132
= 1.15. The errors used in the fit are quadratic combinations of
the statistical and point-to-point systematic errors. As in the
COMPASS case: {\it a}) isospin symmetry for FFs is imposed, {\it
b}) we assume that all unfavored pion FFs are equal and {\it c})
the same parametrizations for the input FFs are used in the
analysis. We find that the description of the proton data (the
mean value of $\chi^2$ per point is equal to 0.96 for $\pi^{+}$
and 0.70 for $\pi^{-}$ multiplicities) is better than that of the
deuteron data (where the mean value of $\chi^2$ per point is equal
to 1.25 for $\pi^{+}$ and 1.31 for $\pi^{-}$ multiplicities). The
quality of the fit to the data is illustrated in Fig. 5 (for the
proton target) and Fig. 6 (for the deuteron target).

Using the extracted FFs from the HERMES data on multiplicities in
the $[Q^2, z]$ presentation we have calculated the multiplicities
at the kinematic points for the $[x, z]$ presentation. The
obtained value for $\chi^2$ is huge, 2093.3 for 144 experimental
points. The results are shown (red curves) in Fig. 3 for a proton
and in Fig. 4 for a deuteron target. As seen from the figures, the
discrepancy is very large for both the proton and deuteron targets
for the first z-bin [0.2-0.3], as well as at lowest x, for all
z-bins. It follows from this observation that the $[x, z]$ and
$[Q^2,z]$ presentation of the HERMES data are not consistent and
that the use of different presentations of the {\it same} data
leads to different physical results. A further study of this
unusual situation is urgently needed.

The extracted pion FFs from the fit to COMPASS data (blue curves)
and from the fit to HERMES data on pion multiplicities (red
curves) are presented in Fig. 7, and compared to those determined
by DSS \cite{DSS} and HKNS (the 2nd reference in \cite{other_FFs})
in Fig. 8. Due to the visible difference in the z region [0.4-0.6]
between the favored fragmentation functions $D_u^{\pi^+}$
extracted from HERMES and COMPASS data, and the large difference
between the corresponding gluon FFs, the blue curves in Fig. 5 and
Fig. 6 corresponding to the multiplicities at the HERMES $[Q^2,z]$
data points calculated by the FFs (COMPASS), lie systematically
lower than the data points for the same z-bin.
\begin{figure} [h]
\begin{center}
  \includegraphics[height=.26\textheight]{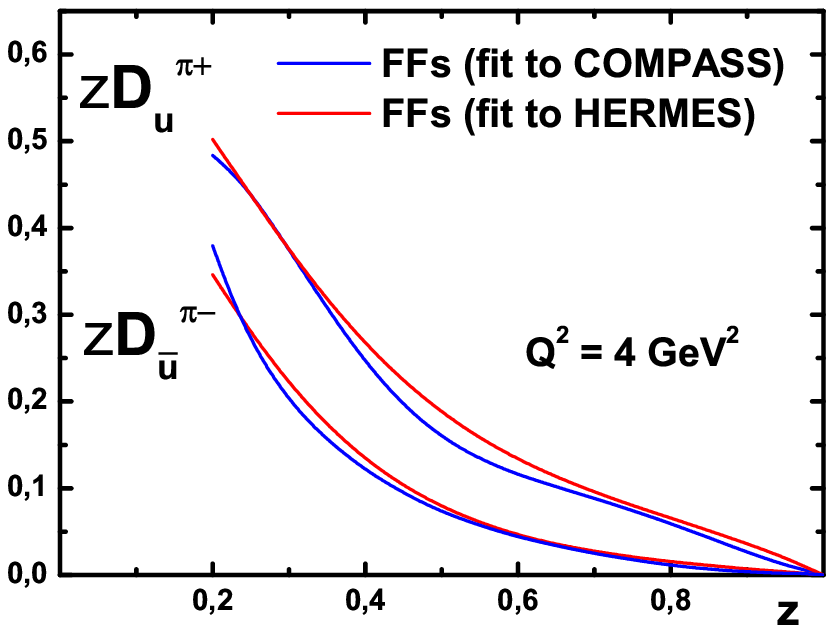}
\includegraphics[height=.26\textheight]{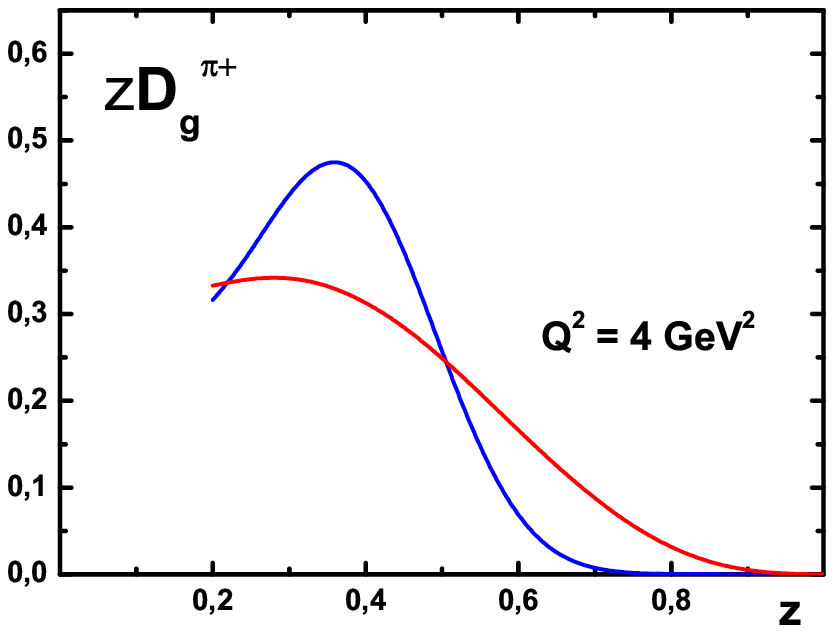}
\caption{\footnotesize Our FFs extracted from the fit to COMPASS
data (blue curves) and HERMES $(Q^2,z)$ data (red curves).}
\label{Sidorov fig17}
\end{center}
\end{figure}
Combined fits to the COMPASS and HERMES $[Q^2, z]$ data on pion
multiplicities will answer the important question if the
discrepancy between the two data sets, shown in Figs. 5 and 6,
will be removed, or more generally, if the HERMES $[Q^2, z]$ and
COMPASS data are or are not consistent.
\begin{figure}
\begin{center}
  \includegraphics[height=.46\textheight]{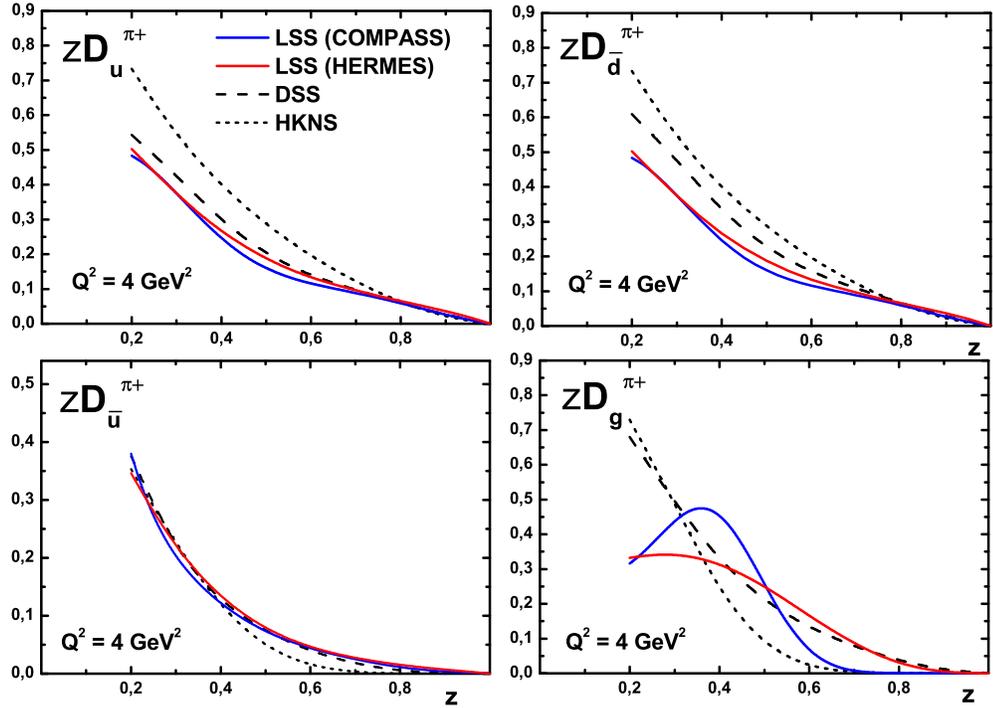}
\caption{\footnotesize Comparison between the new pion FFs and
those of DSS and HKNS} \label{Sidorov fig8}
\end{center}
\end{figure}

One can see from Fig. 8 that the new sets of pion FFs for the
quarks are close to that of DSS. The differences, however, between
$D_g^{\pi^{+}}$ corresponding to the different sets, are large.
Also, for the DSS set the favored fragmentation function $D_{\bar
d}^{\pi^{+}}$ is larger than $D_{u}^{\pi^{+}}$ because in their
analysis a violation of isospin symmetry was allowed. This is the
main reason that the values of the multiplicities calculated by
the DSS FFs for the COMPASS kinematics (blue curves in Figs 1 and
2) are systematically larger then the experimental values. The
situation is the same for the HERMES data.

In conclusion, new sets of pion FFs are determined from the fits
to the recent HERMES and COMPASS data on pion multiplicities. They
differ from those of DSS and HKNS obtained before these data were
available. There is a strong indication that the $[x, z]$ and
$[Q^2,z]$ presentations of the HERMES data on the pion
multiplicities are not equivalent and lead to different physical
results, which suggests that there might be something wrong with
the extraction of the data presentations from the measured
experimental values. We find also that the COMPASS and HERMES $[x,
z]$ data are not consistent. The situation about the consistency
between the COMPASS and HERMES $[Q^2,z]$ data looks much better.
Here the discrepancy is mainly for the third z-bin for the
$\pi^{+}$ and for the second and third z-bins for the $\pi^{-}$
multiplicities. So, the important questions as to the consistency
between COMPASS and HERMES $[Q^2,z]$ data will depend on the
results of a combined fit to the data, which is under way.

\end{document}